\documentclass[prd,reprint,nofootinbib,showpacs,superscriptaddress]{revtex4-1}
\usepackage{graphicx} 
\usepackage{hyperref}
\usepackage{amsfonts}
\usepackage{amsmath,amssymb}
\usepackage{bm} 
\usepackage{color}
\usepackage{epstopdf} 
\usepackage{epsfig}
\usepackage{subfig} 
\usepackage{float}

{\rm }

\def\be{\begin{equation}}
 \def\ee{\end{equation}}
 \def\bea{\begin{eqnarray}}
 \def\eea{\end{eqnarray}}
 \def\bes{\begin{eqnarray}}
 \def\ees{\end{eqnarray}}
 \def\bi{\begin{itemize}}
 \def\ei{\end{itemize}} 

\def\2{\frac{1}{2}}
\def\4{\frac{1}{4}}

\begin{document}

\title{Simultaneous classical communication and quantum key distribution using continuous variable}

\author{Bing Qi}
\email{qib1@ornl.gov}
\affiliation{Quantum Information Science Group, Computational Sciences and Engineering Division,
Oak Ridge National Laboratory, Oak Ridge, TN 37831-6418, USA}
\affiliation{Department of Physics and Astronomy, The
University of Tennessee, Knoxville, TN 37996 - 1200, USA
}

\date{\today}

\begin{abstract}

Presently, classical optical communication systems employing strong laser pulses and quantum key distribution (QKD) systems working at single-photon levels are very different communication modalities. Dedicated devices are commonly required to implement QKD. In this paper, we propose a scheme which allows classical communication and QKD to be implemented simultaneously using the same communication infrastructure. More specially, we propose a coherent communication scheme where both the bits for classical communication and the Gaussian distributed random numbers for QKD are encoded on the \emph{same} weak coherent pulse, and decoded by the \emph{same} coherent receiver. Simulation results based on practical system parameters show that both deterministic classical communication with a bit error rate of $10^{-9}$ and secure key distribution could be achieved over tens of kilometers single mode fibers. It is conceivable that in the future coherent optical communication network, QKD is operated in the background of classical communication at a minimal cost.\footnote{This manuscript has been authored by UT-Battelle, LLC under Contract No. DE-AC05-00OR22725 with the U.S. Department of Energy. The United States Government retains and the publisher, by accepting the article for publication, acknowledges that the United States Government retains a non-exclusive, paid-up, irrevocable, world-wide license to publish or reproduce the published form of this manuscript, or allow others to do so, for United States Government purposes. The Department of Energy will provide public access to these results of federally sponsored research in accordance with the DOE Public Access Plan (http://energy.gov/downloads/doe-public-access-plan).
}

\end{abstract}

\maketitle

\section{Introduction}
\label{sec:1}

One of the most practical applications of quantum information science is the so-called quantum key distribution (QKD) protocol, which allows two remote parties, normally referred to as Alice and Bob, to generate a secure key through an insecure quantum channel \cite{BB84,E91,Gisin02,Scarani09,Lo14,Diamanti16}. The generated key can be further applied in other cryptographic protocols to enhance communication security.

In the first QKD protocol, the Bennett and Brassard 1984 (BB84) QKD protocol \cite{BB84}, information is carried by single-photon signals. Due to the channel loss and other implementation imperfections, those single-photon signals are detected in a probabilistic fashion with a relatively high quantum bit error rate (QBER) in the order of $10^{-2}$ \cite{Zhao06}. This is in sharp contrast to a classical optical communication system, where information can be transmitted deterministically in an almost error-free fashion. Furthermore, specialized devices, such as single photons detectors, are typically required in QKD. This makes QKD a very different communication modality in comparison with classical communication.

More recently, continuous-variable (CV) QKD protocols based on optical coherent detection have been proposed and demonstrated as viable solutions \cite{Ralph99,Hillery00,GMCS,Jouguet13}. In an optical coherent receiver, a strong laser pulse called local oscillator (LO) is mixed with the incoming signal at a beam splitter. The resulting interference signal is strong enough and can be detected using highly efficient photo-diodes working at room temperature. The strong LO also acts as a natural and extremely selective filter, which can effectively suppress broadband noise photons generated in the communication channel. This intrinsic filtering function is especially useful when conducting QKD over a noisy channel, such as a lit fiber in a conventional fiber optic network \cite{Qi10,Kumar15} or a free-space optical link \cite{Heim14}.

One well-known CV-QKD protocol is the Gaussian-modulated coherent states (GMCS) protocol \cite{GMCS}, which has been demonstrated over $100$km telecom fiber \cite{Huang16} and in a real-world CV-QKD network \cite{Huang162}. The hardware required for implementing the GMCS QKD is surprisingly similar to that for classical coherent optical communication \cite{Kikuchi16}. In fact, recent progresses in both communities have significantly reduced the gap between them. On one hand, the continuous improvement of detector performance and the development of forward-error-correction coding in classical coherent communication allows a bit error rate (BER) of $10^{-9}$ with only a few photons per bit \cite{Stevens08}. On the other hand, CV-QKD protocols based on discrete modulation scheme, which are similar to the binary phase-shift keying (BPSK) and the quadrature phase-shift keying (QPSK) in classical communication, have been proposed \cite{Zhao09,Leverrier09}. It is feasible to use the same communication infrastructure for either classical communication or QKD in a time-sharing manner.

In this paper, we go one step further by showing it is possible to conduct classical communication and QKD \emph{simultaneously}. As a specific example, we propose a coherent communication scheme where both the bits for classical communication and the Gaussian distributed random numbers for GMCS QKD are encoded on the \emph{same} weak coherent pulse, and decoded by the \emph{same} coherent receiver. Such a scheme could be appealing in practice, where random numbers for QKD are superimposed on classical communication signals, and secure key distribution is conducted in the background of classical communication at a minimal cost.

Our protocol is a special case of the more general concept of simultaneous transmission of classical and quantum information \cite{Devetak05,Wilde12}. The trade-off relations for public communication, privacy communication and secure key generation have been derived in \cite{Wilde12} assuming no feedback from the receiver to the sender. It could be interesting to extend the results in \cite{Wilde12} to the GMCS QKD using reverse reconciliation.

This work is partly inspired by \cite{Marie16}, where the authors propose improved schemes to implement CV-QKD using a \emph{locally} generated LO \cite{Qi15,Soh15,Huang15}. One scheme proposed in \cite{Marie16} is to estimate the phase difference between two remote lasers by sending Gaussian modulated coherent states displaced in phase space. In our scheme, we encode classical information on the displacements of QKD signals.

This paper is organized as follows: in Section \ref{sec:2}, we present details of the proposed protocol. In Section \ref{sec:3}, we conduct numerical simulations based on practical system parameters. Finally, we conclude this paper with a discussion in Section \ref{sec:4}.
 
\section{Details of the new protocol}
\label{sec:2}

For simplicity, in this paper we assume the BPSK scheme and the GMCS protocol are adopted in classical communication and QKD, correspondingly. The basic idea should be applicable in other encoding schemes.

\subsection{Classical coherent communication using BPSK}

In the BPSK modulation scheme, the classical binary information is encoded on the phase of a coherent state, and decoded by performing optical homodyne detection. More specifically,  the bit value $m_A$ ($m_A=0,1$) is encoded by $|e^{-im_A\pi}\alpha\rangle$, as shown in Fig.1(a). Without loss of generality, we assume $\alpha$ is a real number. The average photon number $\mu$ of the coherent state is given by $\mu=\alpha^2$.

We assume that the X-quadrature of the signal is measured by the receiver. The probability distribution of the measurement result $x_c$ is shown in Fig.1(a) (right figure). The receiver can decode binary information using the sign of $x_c$, i.e., if $x_c>0$ ($<0$), the bit value is assigned as ``0'' (``1''). The BER of the BPSK scheme is determined by the measurement error variance $\sigma$ and the signal amplitude $\alpha$, and is given by \cite{Kikuchi16}
\bes\label{eq1} BER=\dfrac{1}{2}erfc(\dfrac{\sqrt{T_{ch}\eta}\alpha}{\sqrt{2\sigma}})\ees
where $T_{ch}$ is the channel transmittance, $\eta$ is the detection efficiency, and erfc(*) denotes the complementary error function. If the homodyne detector is shot-noise limited (the technical noise is much lower than the vacuum noise), then $\sigma$ is about 1/4. From (1), we need about 9 photons per bit at the receiver's end to achieve a BER of $10^{-9}$.

\begin{figure}[t]
	\includegraphics[width=.45\textwidth]{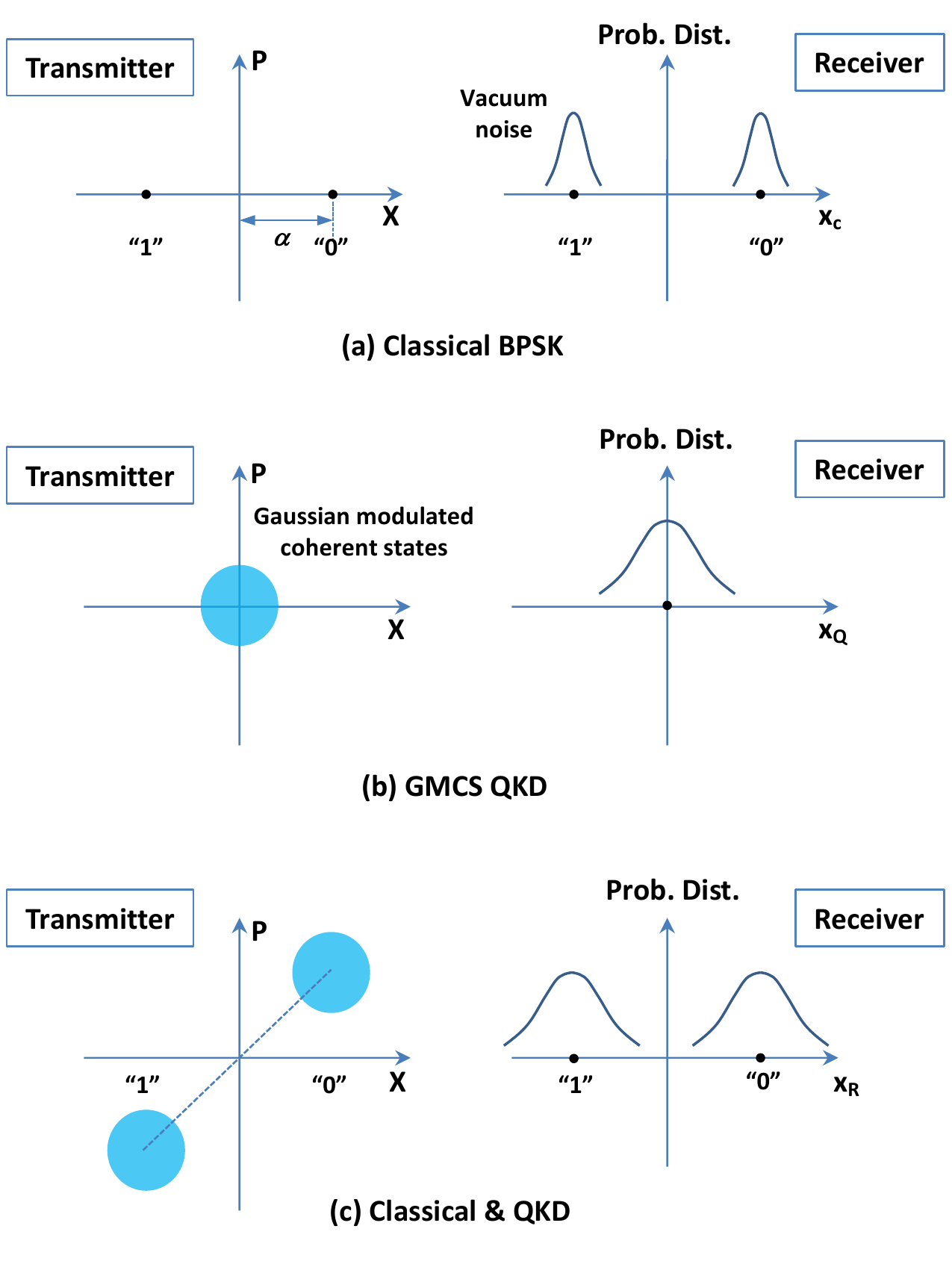}
	\captionsetup{justification=raggedright,
					singlelinecheck=false }
	\caption{Phase space representations of various coherent communication schemes. (a) Classical BPSK scheme. (b) GMCS QKD scheme. (c) The proposed scheme. The figures on the right show the probability distributions of X-quadrature measurement.} 
	\label{fig:1}
\end{figure}

\subsection{GMCS QKD}

In GMCS QKD, Alice prepares coherent states $|x_A+ip_A\rangle$ and sends them to Bob through a quantum channel. Here $x_A$ and $p_A$ are Gaussian random numbers with zero mean and a variance of $V_A N_0$, where $N_0$ = 1/4 denotes the shot-noise variance. In this paper, all the noise variances are defined in the shot-noise unit. At Bob's end, he can either perform optical homodyne detection to measure a randomly chosen quadrature \cite{GMCS}, or perform optical heterodyne detection to measure both quadratures simultaneously \cite{Weedbrook04}. In this paper, our discussion is based on the homodyne detection scheme. The essential ideas can be extended to the other case. Fig.1(b) shows the phase space representation of Gaussian modulated coherent states. Given the system is shot-noise limited, the overall quadrature variance at the receiver's end is $(T_{ch} \eta V_A+1)N_0$. 

After quantum transmission stage, Bob announces which quadrature he measures for each incoming signal through an authenticated classical channel, and Alice only keeps the corresponding data. If the observed noise is below certain threshold, they can further work out a secure key by performing reconciliation and privacy amplification. See more details in Section III.

\subsection{Simultaneous classical communication and QKD protocol}

It is straightforward to combine the above two communication protocols. In this simultaneous classical communication and QKD scheme, Alice encodes her classical bit $m_A$ and Gaussian random numbers $\lbrace x_A, p_A\rbrace$ on a coherent state $|(x_A+e^{-im_A\pi}\alpha)+i(p_A+e^{-im_A\pi}\alpha)\rangle$, as shown in Fig.1(c). Note in the GMCS QKD based on homodyne detection, Bob measures either X or P quadrature of each incoming signal. To achieve deterministic classical communication, the same classical bit $m_A$ is encoded on both X and P quadrature. If heterodyne detection scheme is employed, then Alice only needs to encode $m_A$ on one quadrature.

Assume Bob measures X-quadrature (P-quadrature) and his measurement result is $x_R$ ($p_R$). Bob determines a classical bit $m_B$ using the sign of $x_R$ ($p_R$), i.e., if $x_R (p_R)>0$, then the value of $m_B$ is assigned as ``0''. Otherwise, the value of $m_B$ is assigned as ``1''. To generate secure key, the measurement result will be rescaled and displaced based on the overall transmittance $T_{ch} \eta$ and the value of $m_B$ as follows
\bes\label{eq2} x_B &=& \dfrac{x_R}{\sqrt{T_{ch} \eta}}+(2m_B-1)\alpha \nonumber\\
p_B &=& \dfrac{p_R}{\sqrt{T_{ch} \eta}}+(2m_B-1)\alpha \ees
Alice and Bob can further work out a secure key from raw keys $\lbrace x_A, x_B\rbrace$ and $\lbrace p_A, p_B\rbrace$, just as in the case of conventional GMCS QKD.

Essentially, the states prepared by Alice are displaced Gaussian modulated coherent states, where the amount of displacement is determined by the classical bit $m_A$. Given a modulation variance $V_A$, the BER of classical communication can be reduced effectively by increasing the displacement $\alpha$.

Next, we will show that the security proofs of the standard GMCS QKD can be applied to this new scheme. To illustrate the essential ideas, it is convenient to represent the classical communication and QKD using two separate channels, and allow Eve to have full access and control of information transmitted through the classical channel, as shown in Fig.2 (a). In this picture, Bob performs a homodyne measurement on the incoming signal, then displaces his measurement results using classical information $m_B$ (which could be provided by Eve). Equivalently, Bob could perform the displacement operation first, then perform the homodyne measurement, as shown in Fig.2 (b). Finally, if we move the displacement operation out of Bob's secure station and let Eve to have full control of it (see Fig.2 (c)), then the whole system reduces to the standard GMCS QKD, where Eve is allowed to manipulate the quantum signals transmitted through the channel at her will. Note, in the last step, we have given Eve additional power to control the displacement operation, so the security of the proposed scheme is at least as strong as the standard GMCS QKD.

\begin{figure}[t]
	\includegraphics[width=.4\textwidth]{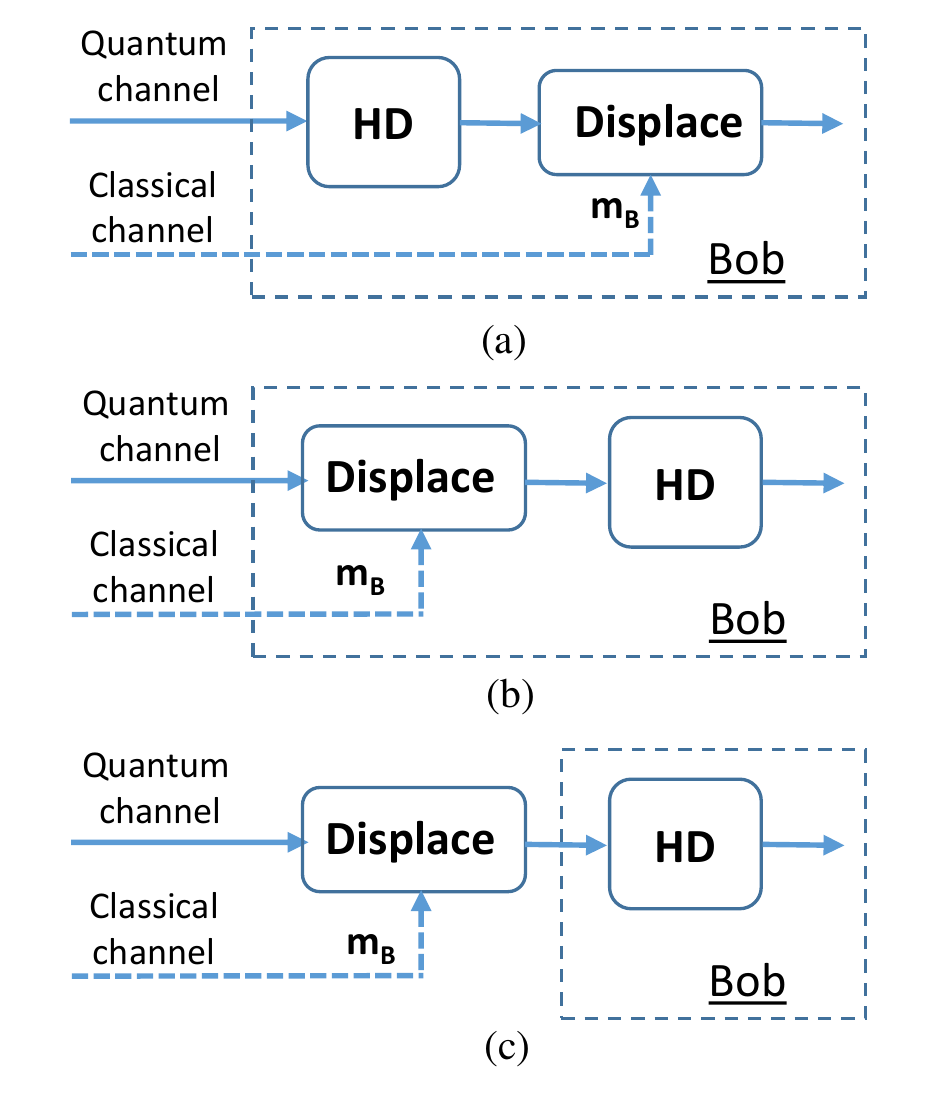}
	\captionsetup{justification=raggedright,
					singlelinecheck=false }
	\caption{Security models. HD--homodyne detection. (a) Our protocol; (b) A virtual QKD scheme equivalents to (a); (c) Standard GMCS QKD. } 
	\label{fig:2}
\end{figure}

In next section, we conduct numerical simulations to estimate the performance of the proposed scheme.

\section{Simulation results}
\label{sec:3}

To evaluate the performance of the proposed scheme, we conduct numerical simulations using practical system parameters. On one hand, since the quantum signal is superimposed on a relatively strong classical signal, we expect that noises from the classical channel will be coupled into the quantum channel and reduce the QKD performance. On the other hand, the Gaussian modulation in QKD will also appear as a noise source in classical communication channel.

We first evaluate the performance of the classical communication. The main noise sources in classical channel are: (1) the vacuum noise with a variance of one; (2) the detector noise denoted by $\nu_{el}$; and (3) the Gaussian modulation for QKD with a variance of $V_A$. All the above noises are defined in the shot-noise unit and are assumed to be independent Gaussian noises with zero mean. Note the first two are defined at the receiver's side, while the third is defined at the sender's side. As we will show below, to achieve a positive key rate in QKD, the excess noise due to phase instability and other modulation imperfections should be much smaller than the vacuum noise. So we neglect the phase noise in classical communication.  

We assume the communication channel is optical fiber with an attenuation coefficient of $\gamma$. The channel transmittance is given by
\bes\label{eq3} T_{ch}=10^{\frac{-\gamma L}{10}}\ees
where $L$ is the fiber length.

From (1), the BER of classical communication is given by
\bes\label{eq4} BER=\dfrac{1}{2}erfc(\dfrac{\sqrt{T_{ch}\eta}\alpha}{\sqrt{2(T_{ch} \eta V_A+1+\nu_{el})N_0}})\ees

To achieve a BER of $10^{-9}$ in the classical channel, the required displacement $\alpha$ is
\bes\label{eq5} \alpha=4.24\times\dfrac{\sqrt{T_{ch} \eta V_A+1+\nu_{el}}}{\sqrt{2T_{ch}\eta}}.\ees

We conduct numerical simulation to calculate the required displacement $\alpha$ as a function of the fiber length using (5). Simulation parameters are $V_A=4$, $\gamma=0.2$ dB/km, $\eta=0.5$, $\nu_{el}=0.1$. The simulation results are shown in Fig.3. 

\begin{figure}[t]
	\includegraphics[width=.5\textwidth]{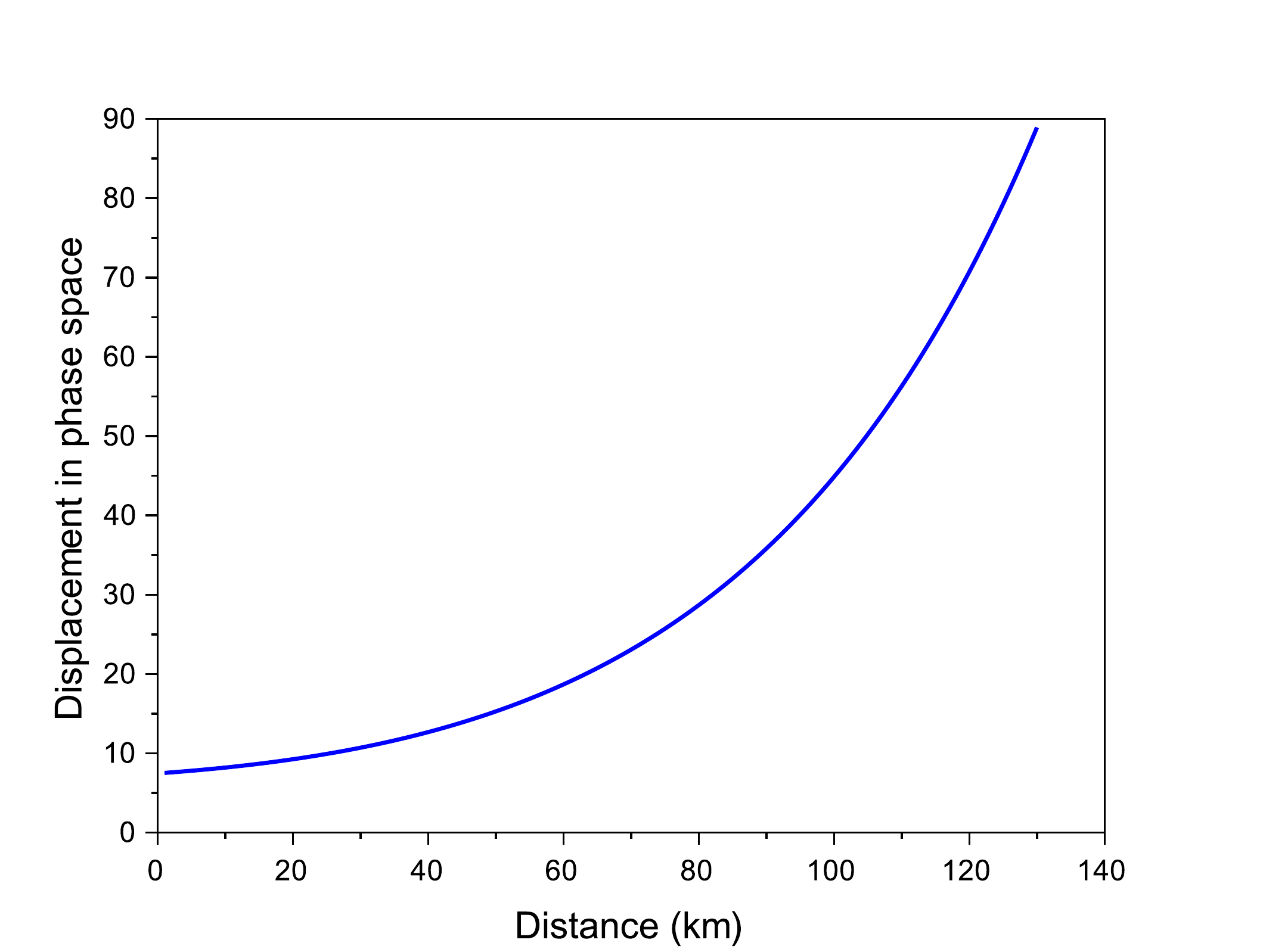}
	\captionsetup{justification=raggedright,
					singlelinecheck=false }
	\caption{Simulation results of the required displacement $\alpha$ to achieve a BER of $10^{-9}$ in the classical channel. Simulation parameters: $V_A=4$, $\gamma=0.2$ dB/km, $\eta=0.5$, $\nu_{el}=0.1$.} 
	\label{fig:3}
\end{figure}

Fig.3 shows that $\alpha^2\geq (V_A+1)N_0$ for a typical $V_A$ in the range of 1 to 20. The presence of this relatively large displacement not only requires a large dynamic range of the detector, but also makes the QKD performance more sensitive to the phase noise. We will study these issues in details below.

The expected quadrature distribution at the receiver's end is shown in Fig.4, where $\alpha'=\sqrt{T_{ch} \eta}\alpha$ is the reduced displacement at receiver's end and $V_B=(T_{ch} \eta V_A+1)N_0$ is the expected variance associated with the Gaussian modulation. Due to the finite dynamic ranges of electrical amplifiers and Analog to Digital Converter (ADC), a practical homodyne detector only has a limited linear response range $[-x_m,x_m]$, within which the output signal is propositional to the quadrature value of the input signal \cite{Chi11, Qin16}. For simplicity, we assume the output of the homodyne detector is either $x_m$ or $-x_m$ if the input signal is beyond the linear range. The excess noise due to the finite measurement range of the detector is denoted by $\varepsilon_c$, which is given by
\begin{multline}
\varepsilon_c=\dfrac{1}{N_0\sqrt{2\pi V_B}} \int_{-\infty}^{-x_m} (x+x_m)^2 e^{-\frac{(x-\alpha')^2}{2V_B}} dx \\ +\dfrac{1}{N_0\sqrt{2\pi V_B}} \int_{x_m}^{\infty} (x-x_m)^2 e^{-\frac{(x-\alpha')^2}{2V_B}} dx.  
\end{multline}

Furthermore, due to the finite resolution of the ADC, we expect a quantization noise variance given by
\begin{equation}
\varepsilon_d=\dfrac{1}{N_0}\left[0.5\times\dfrac{x_m-(-x_m)}{2^M}\right]^2
\end{equation}
where M is the number of bits of the ADC.

To estimate the magnitudes of $\varepsilon_c$ and $\varepsilon_d$ in a practical setup, we assume that the dynamic range of the homodyne detector is 30 dB, corresponding to a practical 10-bit ADC ($M=10$). Other system parameters are assumed to be $V_A=4$, $\gamma=0.2$ dB/km, $\eta=0.5$, and $x_m=10$. Using (5)-(7), $\varepsilon_c$ and $\varepsilon_d$ have been determined to be $4.4\times10^{-9}$ and $3.8\times10^{-4}$, which are negligible comparing with the electrical noise of the detector (typically in the order of $10^{-2}$ to $10^{-1}$). 

\begin{figure}[t]
	\includegraphics[width=.45\textwidth]{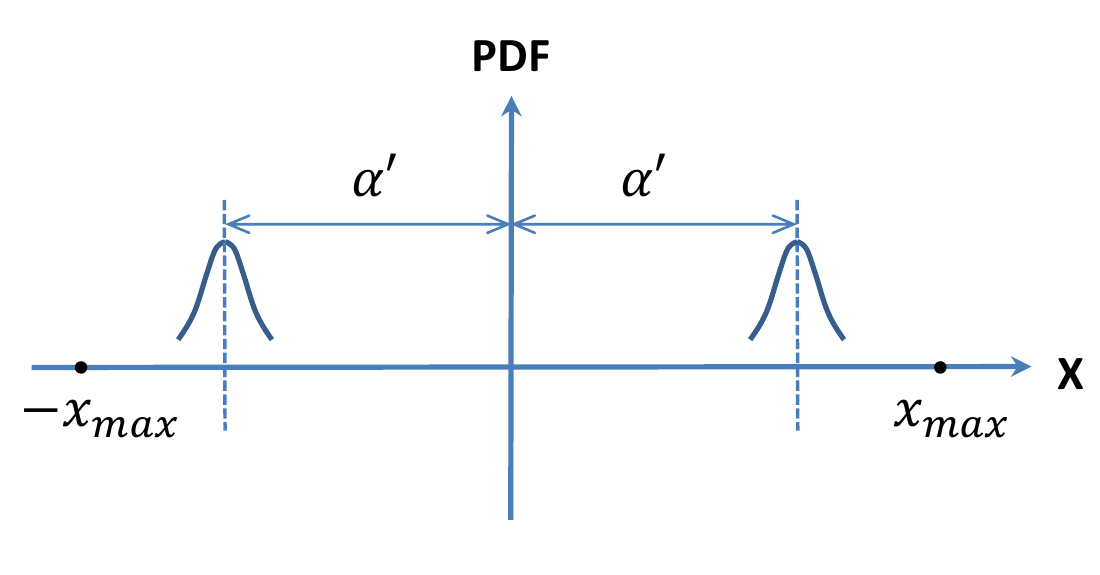}
	\captionsetup{justification=raggedright,
					singlelinecheck=false }
	\caption{The expected quadrature distribution at the receiver's end. $\alpha'=\sqrt{T_{ch} \eta}\alpha$; the variance of the Gaussian distribution is $V_B=(T_{ch} \eta V_A+1)N_0$; the linear range of the homodyne detector is $[-x_m,x_m]$.} 
	\label{fig:4}
\end{figure}

Next, we estimate the impact of phase noise which commonly exists in a coherent communication system. Under the assumption of $\alpha^2\geq (V_A+1)N_0$, the excess noise due to phase instability can be estimated by
\bes\label{eq6} \varepsilon_p=\dfrac{\alpha^2\sigma_\phi}{N_0} \ees
where $\sigma_\phi$ is the phase noise variance. Note $\sigma_\phi$ can include both the phase noise between the signal and the LO, and other modulation errors.

We define the overall excess noise outside of Bob's system as
\bes\label{eq7} \varepsilon=\varepsilon_p+\varepsilon_0 \ees
where $\varepsilon_0$ quantifies the excess noise independent of $\alpha$. For more detailed studies on various noises in CV-QKD, see \cite{Qi07, Lodewyck07, Jouguet12, Marie16}.

To estimate the secure key rate of QKD, we adopt the ``realistic'' model \cite{GMCS} where one crucial assumption is that Eve cannot control the noise and loss inside Bob's system ($\nu_{el}$ and $\eta$). This ``realistic'' model has been widely adopted in long distance CV-QKD experiments \cite{GMCS, Jouguet13, Kumar15, Qi07, Lodewyck07, Huang16}.

The asymptotic secure key rate, under the optimal collective attack, in the case of reverse reconciliation, is given by \cite{Lodewyck07} 
\bes\label{eq8} R=fI_{AB}-\chi_{BE} \ees
where $I_{AB}$ is the Shannon mutual information between Alice and Bob; $f$ is the efficiency of the reconciliation algorithm; $\chi_{BE}$ is the Holevo bound of the information between Eve and Bob. We remark a composable security proof against arbitrary attacks has been developed for GMCS QKD based on the heterodyne detection scheme \cite{Leverrier15}. Unfortunately, the existing security proof cannot produce a positive key rate for a reasonable data size \cite{Diamanti15}. Here, we adopt the security analysis given in \cite{Lodewyck07} to facilitate the comparison with previous experimental results.

The mutual information between Alice and Bob is given by \cite{GMCS}
\bes\label{eq9} I_{AB}=\dfrac{1}{2}log_2\dfrac{V+\chi_{tot}}{1+\chi_{tot}} \ees

The Holevo bound of the information between Eve and Bob is given by \cite{Lodewyck07} 
\bes\label{eq10} \chi_{BE}=\sum_{i=1}^2 G\left( \dfrac{\lambda_i-1}{2} \right)  - \sum_{i=3}^5 G\left( \dfrac{\lambda_i-1}{2}\right)  \ees
where $G(x)=(x+1)log_2(x+1)-xlog_2x$

\bes\label{eq11} \lambda_{1,2}^2=\frac{1}{2} \left[ A\pm \sqrt{A^2-4B} \right] \ees
where
\bes\label{eq12} A=V^2 (1-2T_{ch})+2T_{ch}+T_{ch}^2 (V+\chi_{line})^2 \ees
\bes\label{eq13}B=T_{ch}^2(V\chi_{line}+1)^2 \ees
 
\bes\label{eq14} \lambda_{3,4}^2=\frac{1}{2} \left[ C\pm \sqrt{C^2-4D} \right] \ees
where
\bes\label{eq15}C=\dfrac{A\chi_{hom}+V\sqrt{B}+T_{ch}(V+\chi_{line})}{T_{ch}(V+\chi_{tot})} \ees 
\bes\label{eq16}D=\sqrt{B}\dfrac{V+\sqrt{B}\chi_{hom}}{T_{ch}(V+\chi_{tot})} \ees 
    
\bes\label{eq17} \lambda_5=1 \ees

\begin{figure}[t]
	\includegraphics[width=.5\textwidth]{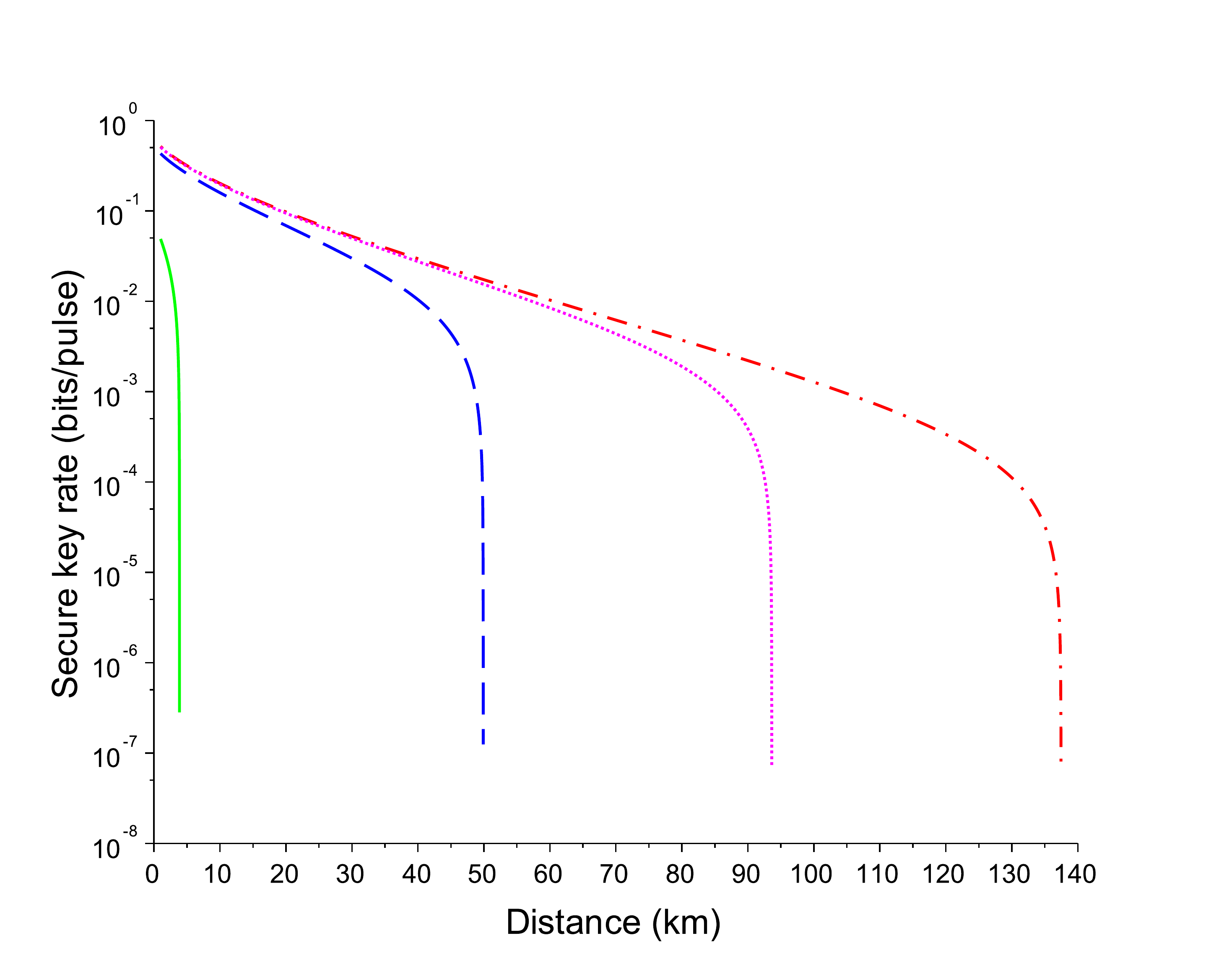}
	\captionsetup{justification=raggedright,
					singlelinecheck=false }
	\caption{Simulation results based on security analysis given in \cite{Lodewyck07}. Simulation parameters: $\gamma=0.2$ dB/km; $\varepsilon_0=0.01$; $\nu_{el}=0.1$; $V_A=4$; $\eta=0.5$; $f=0.95$; and $\sigma_\phi$=$10^{-3}$ (solid line), $10^{-4}$ (dash line), $10^{-5}$ (dot line) and $10^{-6}$ (dash dot line).} 
	\label{fig:5}
\end{figure}

System parameters in the above equations are defined as follows:
\begin{itemize}
\item[{\em (a)}] $V=V_A+1$.
\item[{\em (b)}] The total noise referred to the channel input $\chi_{tot}=\chi_{line}+\dfrac{\chi_{hom}}{T_{ch}}$.
\item[{\em (c)}] The total channel-added noise referred to the channel input $\chi_{line}=\frac{1}{T_{ch}}-1+\varepsilon$. 
\item[{\em (d)}] The detector-added noise referred to Bob's input $\chi_{hom}=[1-\eta+\nu_{el}]/\eta$.
\end{itemize}

We conduct numerical simulations of secure key rate at different phase noise variances. Other simulation parameters are: $\gamma=0.2$ dB/km, $\varepsilon_0=0.01$, $\nu_{el}=0.1$, $\eta=0.5$, $f=0.95$, and $V_A=4$. Fig.~\ref{fig:5} shows the simulation results. The performance of the proposed scheme is heavily dependent on the phase noise of the system. The observed phase noises in previous CV-QKD experiments are in the orders of $10^{-3}$ \cite{Qi07}, $10^{-4}$ \cite{Lodewyck07}, and $10^{-6}$ \cite{Huang16}, suggesting simultaneous classical communication and QKD over tens of kilometer optical fiber is possible. 

\section{Discussion}
\label{sec:4}

One major roadblock to the wide applications of QKD is the high implementation cost. Presently, dedicated hardware are required to implement QKD protocols. In this paper, we show that by using optical coherent detection, classical communication and QKD could be implemented simultaneously on the same platform. Our simulation results suggest that the QKD performance is largely determined by the phase noise of the coherent communication system. To extend the distance of QKD, sophisticate phase stabilization scheme may be required.

In our simulation, an uncoded BPSK scheme is assumed for classical communication. In modern optical communication systems, forward error correction (FEC) schemes are commonly used to achieve high accuracy of data transmission with the minimum of optical power \cite{Chang10}. For example, by applying FEC coding, a BER of $10^{-9}$ was achieved at a signal power of 1.5 photons per pulse \cite{Stevens08}. In comparison, an uncoded BPSK system requires at least 9 photons per pulse to achieve the same BER. As shown in (8), the excess noise due to the phase instability is proportional to the optical power in classical channel. By applying FEC coding, the excess noise in QKD can be further reduced. This can lead to a better secure key rate or a longer distance.

To apply our scheme in practice, there are important challenges to be addressed.

First, in our simulation, the secure key rate is calculated based on a security proof against collective attacks in the asymptotic limit of infinitely large data size. In practical applications, a composable security against arbitrary attacks is required. Unfortunately, the existing composable security proof techniques cannot produce a positive key rate for a reasonable data size \cite{Diamanti15}. This is a significant challenge not only to our scheme but also to all CV-QKD protocols based on coherent states. As remarked by the authors of \cite{Diamanti15}, finding better security proofs allowing composable security for reasonable data size ``is certainly the most pressing issue in the theoretical study of CVQKD''.

Second, while today's classical communication systems can be operated at 10 to 100 GHz repetition rate, the highest repetition rate demonstrated in CV-QKD experiments is only 100 MHz \cite{Huang15}. To implement the simultaneous classical and quantum communication protocol, we either have to sacrifice the speed of the classical communication, or need to develop high-speed (above 10 GHz) shot-noise limited homodyne detectors. We remark that the speed of shot-noise limited homodyne detectors has been improved significantly over the years. A 1 GHZ shot-noise limited homodyne detector has been applied in a recently CV-QKD experiment \cite{Huang15}. We expect this trend will continue in the future and the gap between the repetition rates of classical and quantum communications may eventually disappear.

We acknowledge helpful comments from Romain All\'eaume and Mark Wilde. This work was performed at Oak Ridge National Laboratory (ORNL), operated by UT-Battelle for the U.S. Department of Energy under Contract No. DE-AC05-00OR22725. The authors acknowledge support from ORNL laboratory directed research and development program.

\end{document}